\documentclass[prb,twocolumn,showpacs,superscriptaddress, amsmath,amssymb]{revtex4}

\usepackage{graphicx}
\usepackage{dcolumn}
\usepackage{bm}
\usepackage[dvips]{color}
\usepackage{color}

\newcommand{\vect}[1]{\mathbf{#1}}

\begin{document}
\preprint{APS/123-QED}

\title{Search for a nematic phase in quasi-2D antiferromagnet CuCrO$_2$
by NMR in electric field}

\author{Yu.A. Sakhratov}
\affiliation{National High Magnetic Field Laboratory, Tallahassee, Florida
32310, USA} \affiliation{Kazan State Power Engineering University, Kazan 420066, Russia}

\author{J.J. Kweon}
\affiliation{National High Magnetic Field Laboratory, Tallahassee, Florida
32310, USA}

\author{E.S. Choi}
\affiliation{National High Magnetic Field Laboratory, Tallahassee, Florida
32310, USA}

\author{H.D. Zhou}
\affiliation{National High Magnetic Field Laboratory, Tallahassee, Florida
32310, USA} \affiliation{Department of Physics and Astronomy, University of
Tennessee, Knoxville, Tennessee 37996, USA}

\author{L.E. Svistov}
\email{svistov@kapitza.ras.ru}
\affiliation{P.L. Kapitza Institute for Physical Problems, RAS, Moscow 119334 , Russia}

\author{A.P.~Reyes}
\affiliation{National High Magnetic Field Laboratory, Tallahassee, Florida
32310, USA}

\date{\today}

\begin{abstract}
The magnetic phase diagram of CuCrO$_2$ was studied with a novel method of simultaneous Cu NMR and electric polarization techniques with the primary goal of demonstrating that regardless of cooling history of the sample the magnetic phase with specific helmet-shaped NMR spectra associated with interplanar disorder possesses electric polarization.
Our result unequivocally confirms the assumption of Sakhratov \emph{et al.} [Phys. Rev. B {\bf{94}}, 094410 (2016)]
that the high-field low-temperature phase is in fact a 3D-polar phase characterised by a 3D magnetic order with tensor order parameter. In comparison with the results obtained in pulsed fields, a modified phase diagram is introduced defining the upper boundary of the first-order transition from the 3D-spiral to the 3D-polar phase.
\end{abstract}


\maketitle

\section{Introduction}
CuCrO$_2$ is an example of quasi-two-dimensional (quasi-2D) antiferromagnet ($S=3/2$) with triangular lattice structure.
Below $T_c\approx 24$~K CuCrO$_2$ orders in three-dimensional (3D) planar spiral structure with incommensurate wave vector directed along one side of the triangle. In the ordered phase spontaneous electric polarization directed perpendicular to the spin plane appears.\cite{Zapf_2014}

In this system exotic types of magnetic orderings are expected due to geometrical frustrations of in-plane and interplanar exchange interactions. Electric polarization studies of CuCrO$_2$ in pulsed magnetic fields up to 92~T\cite{Zapf_2014, Lin_2014} revealed a rich phase diagram. The nature of realized magnetic phases were analyzed from theoretical consideration of exchange bonds in Ref.~[\onlinecite{Lin_2014}]. In recent Cu NMR investigations\cite{Sakhratov_2014, Sakhratov_2016} the hysteretic behaviour of spectral shapes in the field range 14-45~T which is much below the fields of exchange scale ($\mu_0H_{sat}\approx 280$~T) was observed. The field evolution of NMR spectra was explained by a loss of usual interplanar ordering in the high field range.  However, at the same fields and temperatures CuCrO$_2$ demonstrates multiferroic properties,\cite{Zapf_2014} which indicates the 3D magnetic order. In order to reconcile NMR and electric polarization observations, it was suggested\cite{Sakhratov_2016} that the high field  phase is, in fact, a 3D-polar phase with tensor order parameter which can be classified as a $p$-type nematic phase.\cite{Andreev_1984}

In the present work we continue NMR studies of the polar phase. For this purpose an NMR probe
was modified to allow simultaneous measurements of NMR spectra and electric polarization.
From the NMR point of view, it is verified that the disordered phase does possess spontaneous electric polarization. Furthermore the boundary of the transition from the 3D spiral phase to the high-field 3D-polar phase is investigated.

\section{Crystal and magnetic structure}
\begin{figure}
\includegraphics[width=0.7\columnwidth,angle=0,clip]{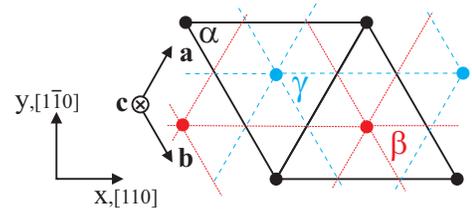}
\caption{(color online)
Crystal structure of CuCrO$_2$ projected on the $ab$-plane.
The three layers, $\alpha\beta\gamma$, are the positions of Cr$^{3+}$ ions.
}
\label{fig:fig1}
\end{figure}

The structure CuCrO$_2$ consists of magnetic Cr$^{3+}$ (3$d^3$, $S=3/2$),
nonmagnetic Cu$^+$, and O$^{2-}$ triangular lattice planes, which are stacked along $c$-axis in the sequence Cr-O-Cu-O-Cr
(space group $R\bar{3}m$). The layer stacking sequences are $\alpha\gamma\beta$, $\beta\alpha\gamma$,
and $\beta\beta\alpha\alpha\gamma\gamma$ for Cr, Cu and O ions, respectively.
The crystal structure of CuCrO$_2$ projected on the $ab$-plane is shown in Fig.1.
The magnetic ordering in CuCrO$_2$ occurs in two stages.\cite{Frontzek_2012}
At the higher transition temperature $T_{c1}=24.2$~K, a transition to a 2D ordered state occurs,
whereas below $T_{c2}=23.6$~K, a 3D magnetic order with incommensurate propagation vector
$\vect{k}_{ic}= (0.329,0.329,0)$ is established. The magnetic ordering is accompanied by a weak crystallographic distortion of one side of the triangle which is parallel to $\vect{k}_{ic}$.\cite{Kimura_JPSJ_2009}
The magnetic moments of Cr$^{3+}$ ions in zero magnetic field can be described by the expression
\begin{eqnarray}
\vect{M}(\vect{r}_{i,j})=M_1\vect{e}_1\cos(\vect{k}_{ic}\vect{r}_{i,j}+\Theta)+M_2\vect{e}_2\sin(\vect{k}_{ic}\vect{r}_{i,j}+\Theta),
\label{eqn:spiral}
\end{eqnarray}
where $\vect{e}_1$ and $\vect{e}_2$ are two perpendicular unit vectors
determining the spin plane orientation with the normal vector
$\vect{n}=\vect{e}_1 \times \vect{e}_2$, $\vect{r}_{i,j}$ is the vector to the
$i,j$-th magnetic ion and $\Theta$ is an arbitrary phase. For zero magnetic field
$\vect{e}_1$ is parallel to $[001]$ with $M_1 = 2.8(2)~\mu_B$, while
$\vect{e}_2$ is parallel to $[1\bar{1}0]$ with $M_2 = 2.2(2)~\mu_B$.~\cite{Frontzek_2012}
Below $T_{c2}$ a non zero polarization $\vect{P}$ appears and its direction is defined by the direction of the vector $\vect{n}$.\cite{Seki_2008,Kimura_2008}

The orientation of the spin plane is defined by the biaxial crystal anisotropy.
One {\it hard} axis for the normal vector $\vect{n}$ is perpendicular to the triangular planes and the second axis lies in $ab$-plane perpendicular to the distorted side of the triangle. The
anisotropy along $c$ direction dominates with the anisotropy constant approximately
hundred times larger than that within $ab$-plane. A magnetic phase transition was observed for the
field applied perpendicular to one side of the triangle ($\vect{H}\parallel
[1\bar{1}0]$) at $\mu_0H_c = 5.3$~T, which was consistently
described~\cite{Kimura_PRL_2009, Soda_2010, Svistov_2013} by the reorientation
of the spin plane from $(110)$ ($\vect{n}\perp\vect{H}$) to $(1\bar{1}0)$
$(\vect{n}\parallel\vect{H})$. This spin reorientation to ``umbrella like'' phase
happens due to the weak susceptibility anisotropy of the spin structure
$\chi_{\parallel}\approx 1.05\chi_{\perp}$, where $\parallel$ and $\perp$
refer to fields parallel and perpendicular to $\vect{n}$, respectively.

Due to the symmetry there are six possible directions for the vector $\vect{n}$, so there are
six magnetoelectric domains. In the absence of both electric and magnetic fields these domains are degenerate.
According to Refs.~[\onlinecite{Kimura_2008, Kimura_PRL_2009, Soda_2010, Svistov_2013, Sakhratov_2014}], the application of external electric field $\vect{E}$ along $[110]$ favors the three domains with positive projections of ${n}_{E}$,
while the application of external magnetic field along $[110]$ favors the two domains with $\vect{n}$ collinear to $\vect{H}$. This scheme allows control over the domain arrangement of a sample. For magnetic field $\vect{H} \parallel \vect{c}$, i.e. aligned along the strong anisotropy axis, all six domains are expected to be equivalent.

Magnetic field directed in $ab$-plane acts only on the spin plane orientation and does not change the spin structure at least up to fields 92~T.\cite{Zapf_2014, Lin_2014} For fields directed parallel to the $c$ axis a number of phase transitions were observed.\cite{Zapf_2014, Lin_2014} The transition observed at $\approx 15$~T was interpreted in Ref.~[\onlinecite{Sakhratov_2016}] as a transition to 3D ordered phase with tensor order parameter.
The main properties of magnetic structure realized in CuCrO$_2$ at low magnetic fields such as incommensurability, the peculiarities of magnetic anisotropy, electrical polarization and weak elastic distortions of the lattice can be naturally explained in the frame of Dzyaloshinskii-Landau theory of magnetic phase transitions given in Ref.~[\onlinecite{Marchenko_2014}].

\section{Sample preparation and experimental details}
\begin{figure}
\includegraphics[width=0.6\columnwidth,angle=0,clip]{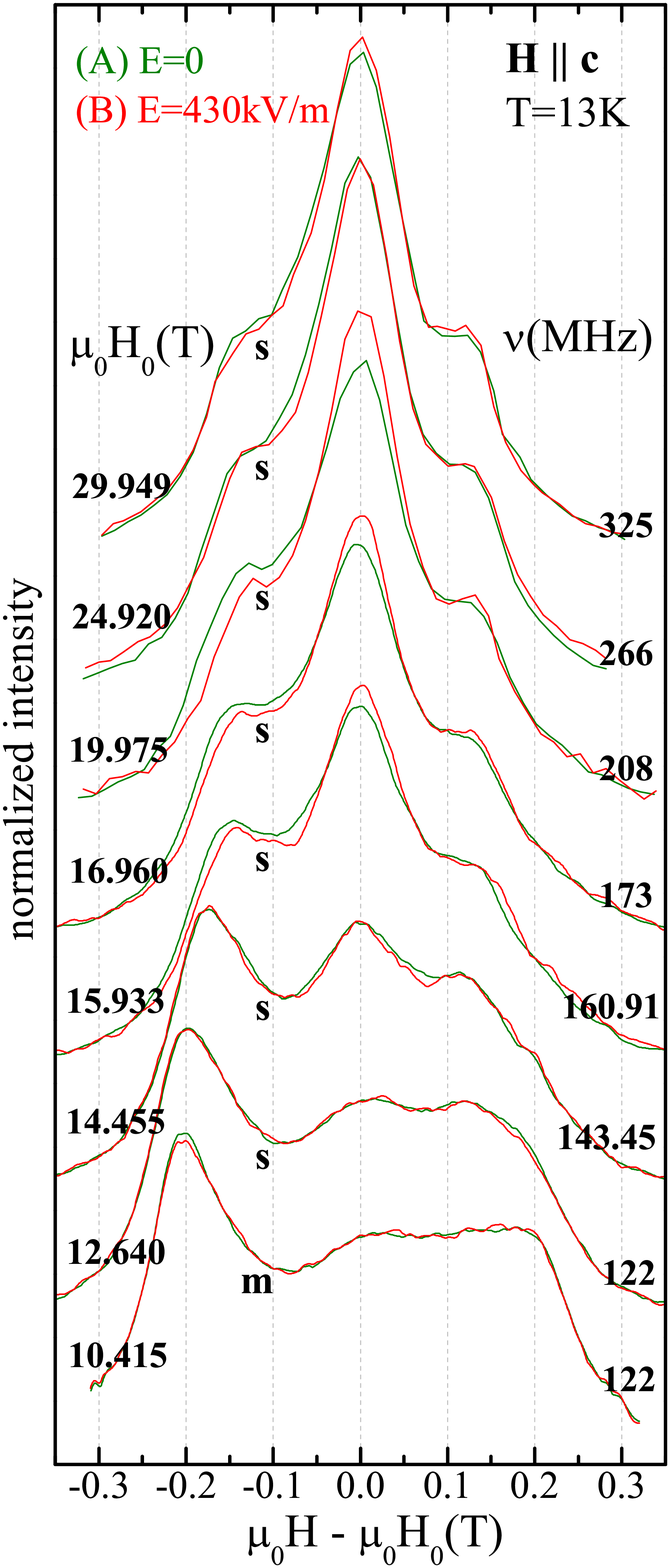}
\caption{(color online)
Field evolutions of $^{63}$Cu NMR spectra measured at temperature 13~K and $\vect{H} \parallel \vect{c}$ after zero magnetic field cooling.
The electric field $E=0$ during cooling for the green spectra (A) and $E=430$~kV for the red spectra (B).
The spectra are shifted by the values $\mu_0H_0$ indicated to the left of each line. Symbols $m$ and $s$ correspond to the main ($m_I = +1/2\leftrightarrow -1/2$) and high-field satellite ($m_I = +3/2\leftrightarrow +1/2$) transitions, respectively. The spectra are taken successively in increasing field.
}
\label{fig:fig2}
\end{figure}

A single crystal of CuCrO$_2$ was grown by the method described in our previous work.\cite{Sakhratov_2014} The crystal was cut into a thin plate with dimensions $2.4\times1.5\times0.7$~mm$^3$. Silver epoxy was painted on the widest end faces and served as the electrodes. The smallest dimension corresponds to the $[110]$ direction of the crystal. The NMR coil is oriented such that the RF field is parallel to the electrode plates. Applying a constant voltage of 300~V to the electrodes corresponds to a poling electric field $E\sim430$~kV/m in $[110]$ direction.

NMR measurements were taken on a superconducting Cryomagnetics 17.5~T magnet and a 30~T resistive magnet at the National High Magnetic Field Laboratory. All magnets were field sweepable. The construction of NMR probe allowed measuring NMR in the presence of the poling electric field. $^{63}$Cu nuclei (nuclear spin $I=3/2$, gyromagnetic
ratio $\gamma/2\pi=11.285$~MHz/T) were probed using pulsed NMR technique.
The spectra were obtained by summing fast Fourier transforms (FFT)
or integrating the averaged spin-echo signals as the field was swept through the resonance line.
NMR spin echoes were obtained using $\tau_p~-~\tau_D~-~2\tau_p$ pulse sequences, where the pulse lengths $\tau_p$ were 1.1-1.7~$\mu$s and the times between pulses $\tau_D$ were 15~$\mu$s.
Measurements were carried out in the temperature range $5\leq T \leq 30$~K stabilized with a
precision better than 0.1~K.

To obtain temperature dependent part of the electric polarization, the pyroelectric current was measured and then integrated with time. The rate of the temperature sweep during warming up was 40~K/min.

Both the NMR and the polarization measurements were taken under different cooling conditions, in which the sample was subjected to various combinations of magnetic and electric fields. For the case of electric field cooled experiment, the electric field remained on during the NMR measurements and then turned off before the pyroelectric current is measured.

\section{Experimental results and discussion}

\begin{figure}
\includegraphics[width=0.9\columnwidth,angle=0,clip]{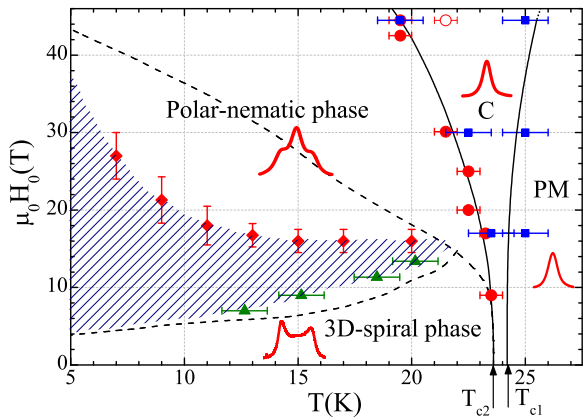}
\caption{(color online)
H-T magnetic phase diagram of CuCrO$_2$ for $\vect{H}\parallel \vect{c}$. The dashed lines are from Ref.\onlinecite{Zapf_2014}. The red diamonds are obtained in the present work and the rest are from Ref.\onlinecite{Sakhratov_2016}. The hysteresis area is shaded. Typical NMR spectral shapes associated with each phase are illustrated.
}
\label{fig:fig3}
\end{figure}

In this paper we investigate the phase transition from a 3D-ordered spiral state to a 3D ordered state with tensor order parameter.\cite{Sakhratov_2016} Fig.2 shows field evolutions of the spectra with increasing field after zero magnetic field cooling from 30~K to 13~K. At low fields the spectra have distorted double-horn shape, at high fields the spectra are helmet-shaped. The observed spectral shapes reflect different magnetic structures realized in external magnetic fields. As elaborated in Refs.[\onlinecite{Sakhratov_2014,Sakhratov_2016}] the ``double-horn'' and ``helmet'' shaped NMR spectra in CuCrO$_2$ are specific for incommensurate magnetic structures within triangular planes with  inter-plane order and disorder, respectively. We emphasize that the high-field helmet shaped spectra persists up to 45~T (Ref.[\onlinecite{Sakhratov_2016}]). Similar measurements carried out at different temperatures allow us to determine the upper boundary of the first-order phase transition from the 3D spiral to the 3D phase with tensor order parameter. The obtained points of phase transitions are shown in Fig.3 with red diamonds. For $T = 5$~K the transition was above 30~T which exceeded our experimental capability. The rest of the points are taken from Ref.[\onlinecite{Sakhratov_2016}].
There are two regions, marked as '3D-spiral phase' and 'Polar nematic phase', that are identified with either double-horn or helmet-shaped spectra. These regions are separated by a broad region, shown shaded, where the spectra demonstrate hysteretic behavior. The hysteresis region is substantially smaller than the transition region observed in Refs.~[\onlinecite{Zapf_2014, Lin_2014}]. This may be due to the fact that the measurements\cite{Zapf_2014, Lin_2014} were carried out in pulsed fields or that the samples were of different quality. Two regions, identified as C and PM, demonstrate single peaked NMR spectra and can be attributed to a collinear up-up-down (UUD) and paramagnetic structures, respectively.\cite{Sakhratov_2016}

The electric field does not appreciably change the shape of the spectra in Fig.2. Nevertheless note that near the transition (15.933~T $\leq \mu_0H_0 \leq$ 19.975~T) there is a difference between the $E=0$ spectra (A) and the $E\neq0$ spectra (B): The intensity of the central peak as compared to the left and right ones is bigger for the spectra (B). This difference is well reproduced at different temperatures. It means that in the case (B) the helmet-shaped state occurs slightly earlier on increasing field. The origin of this anomaly is still unknown and is a subject of a future study.

According to Ref.[\onlinecite{Zapf_2014}], helmet shaped spectra region possesses polarization.
To ensure that this polarization is not related to the fact that the magnetic structure has not enough time to rearrange itself during pulsed field sweep,\cite{Zapf_2014} the measurements of the pyroelectric current in constant magnetic field are performed. Other kinds of cooling conditions that may lead to any new spectral shapes are also checked.

\begin{figure}
\includegraphics[width=0.9\columnwidth,angle=0,clip]{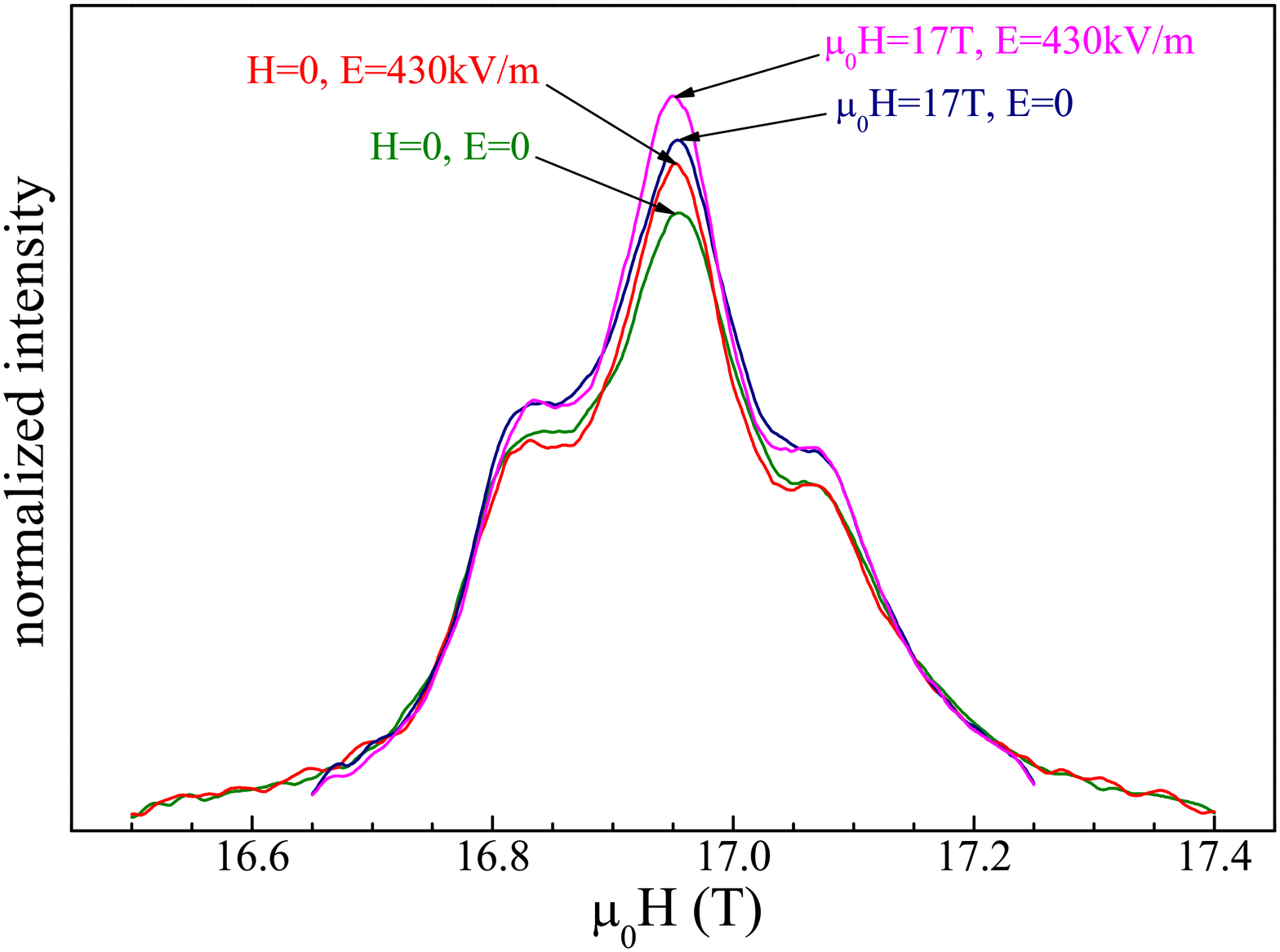}
\caption{(color online)
$^{63}$Cu NMR spectra ($m_I = +3/2\leftrightarrow +1/2$ transition), $\vect{H}\parallel \vect{c}$, $\nu = 173$~MHz, $T = 15$~K. The sample was cooled from $T = 30$~K in the fields $\vect{H}\parallel \vect{c}$, $\vect{E}\parallel [110]$ with the  values given in the figure.
}
\label{fig:fig4}
\end{figure}

\begin{figure}
\includegraphics[width=0.9\columnwidth,angle=0,clip]{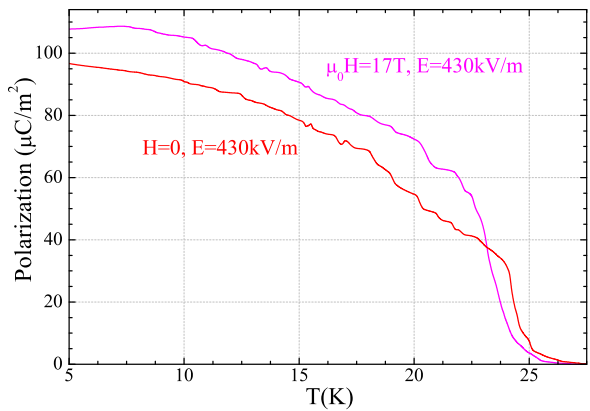}
\caption{(color online)
Temperature dependence of polarization along [110]. The sample was cooled from $T = 30$~K in the fields $\vect{H}\parallel \vect{c}$, $\vect{E}\parallel [110]$ with the  values given in the figure.
}
\label{fig:fig5}
\end{figure}

Fig.4 shows an example of $^{63}$Cu NMR spectra measured at the temperature of 15~K after cooling from 30~K in various conditions specified in the figure. We measured the spectra of the sample cooled in these ways at different temperatures and found no new spectral shapes. The spectra remain helmet shaped which indicates the absence of conventional long-range order. In addition to NMR the pyroelectric current in the same sample after the same cooling conditions was measured. Fig.5 shows the temperature dependence of the polarization. The obtained values of polarization agree well with the values in previous studies\cite{Zapf_2014, Kimura_2008} with slight increase of polarization at 17~T. Thus it was proven that helmet-shaped phase indeed possesses polarization and the assumption of its tensor order parameter\cite{Sakhratov_2016} seems realistic.

\section{Conclusions}
The magnetic phase diagram of CuCrO$_2$ was studied by $^{63}$Cu NMR simultaneous with electric polarization techniques. It has been proven that the magnetic phase with specific helmet-shaped NMR spectra associated with interplanar disorder possesses electric polarization regardless of cooling history of the sample. This result confirms the assumption of Ref.[\onlinecite{Sakhratov_2016}] about the 3D magnetic order with tensor order parameter within this phase which can be classified as a polar-nematic phase.
Note also that NMR technique gives information about the distribution of the local fields on the nuclei, i.e. it cannot give information about the correlation length within individual triangular plane in the nematic phase. These aspects make CuCrO$_2$ intriguing for future experimental and theoretical studies.

\acknowledgments
We thank V.I.~Marchenko for stimulating discussions.
H.D.Z. thanks for the support from  NSF-DMR through award DMR-1350002.
The work was supported by Russian Foundation for Basic Research (Grant 16-02-00688).

Work at the National High Magnetic Field Laboratory is
supported by the NSF Cooperative Agreement No.~DMR-1157490, the State of
Florida, and the DOE.

\end{document}